\documentstyle[preprint,aps]{revtex}
\begin{document}
\thispagestyle{empty}
\title{Flavor Asymmetry in Hyperons and Drell-Yan Processes
\thanks{This report is based on work carried out with M. Alberg, T. Falter,
X. Ji, and A.W. Thomas}}

\author{E.M.Henley \thanks{Supported in part by the DOE}}
\address{Department of Physics and Institute for Nuclear Theory
,University of Washington,\\ 
Box 351560, Seattle, WA 98195, USA }
\maketitle
\bibliographystyle{unsrt}
\date{today}

\begin{abstract}
SU(3), baryon octets and a meson cloud model are compared for the flavor
asymmetry of sea quarks in the $\Sigma^+$, as an example. Large differences 
are found, especially between SU(3) and the meson cloud model. We suggest
Drell-Yan measurements of $\Sigma^{+}-p$ and $\Sigma^{+}- d$ to test the 
prediction
of various models. We use the meson cloud model to predict both valence and
sea quark distributions. 
\end{abstract} 
\newpage

    The predicted \cite {Thom} and measured
\cite{NA51},\cite{E866} flavor asymmetry of the
 proton, e.g., 
$\bar{d}$ / $\bar{u}$ or$ (\bar{d} - \bar{u})$ have awakened considerable 
interest.
A simple explanation first proposed by A. Thomas \cite {Thom} was in terms of
 the pion sea surrounding the quarks in the proton. Since a proton consists of
$uud$ quarks surrounded by a $\pi^0$ or $udd$ quarks surrounded by a 
$\pi^+ (u\bar{d})$
, an excess of $\bar{d}$ over $\bar{u}$ is to be expected. This model has been 
examined quantitatively \cite {Sp} and can explain the Gottfried sum rule
deficiency \cite{NMC} and the ($\bar{d} - \bar{u}$) measured in Drell-Yan p -p 
collisions \cite{NA51},\cite{E866}
    We have examined the expected flavor asymmetry of $\Sigma ^\pm$ baryons.
\cite {Alb} We find large differences between the expected asymmetry on 
(1) the basis of SU(3), (2) a baryon octet $\otimes $meson octet model, and
(3) the meson cloud model. Thus, a measurement of the sea asymmetry 
via the Drell-Yan reactions $\Sigma^\pm + p and \Sigma^pm + d
\rightarrow l^+ +  l^- + X
$ can be used to differentiate between the models.

    I will use the $\Sigma^+$ to illustrate the thesis. It is very easy to
understand the difference between SU(3) and the meson cloud model. In SU(3)
we have
$    u \bar{u}$ in the $p  \Rightarrow  u \bar{u}$ in the $\Sigma^+$,
   $ d \bar{d}$ in the $p  \Rightarrow  s \bar{s}$ in the $\Sigma^+$,
   $ s \bar{s}$ in the $p  \Rightarrow  d \bar{d}$ in the $\Sigma^+$.
Thus, we have $\frac{\bar{d}}{\bar{u}}(\Sigma^+) = \frac {\bar{s}}{\bar{u}}
(p)$ in SU(3). On the other hand, one expects a larger $\bar{d}/ \bar{u}$
ratio in the $\Sigma^+$
than in the proton in the meson cloud model because $\Sigma^+$ can decompose 
into $\Sigma^+ \pi^0, \Sigma^0 \pi^+, \Lambda^0 \pi^+,
$ and $p \bar{K}^0$, where all
but the first case ($\Sigma^+ \pi^0$) correspond to an excess of 
$\bar{d}$ quarks.

    At $x\sim 0.2$, the measured ratio $\bar{u}/\bar{d} \approx 1/2$
\cite{NA51}
 (or 2/3 \cite{E866}); also $\frac{\bar{s}}{\bar{u} + \bar{d}} (p)
\approx 1/4$. This gives $\frac{\bar{d}}{\bar{u}}(\Sigma^+) = \frac {\bar{s}}
{\bar{u}}(p) \sim 0.7$ in SU(3); this value is $< 1$, in contrast to the meson
cloud model. 

    We have also examined a proton made up of a baryon octet $\otimes$ a 
meson octet with a ratio of SU(3) couplings  F/D = 0.6. A summary is presented
in table I for $x \simeq 0.2$.

\begin{table} 
\caption{ Predicted and measured flavor ratios. The experimental column
refers to the proton; all other ones are predictions for the $\Sigma^+$.}
\begin{tabular}{l|rrrr}
Flavor ratios &Experim. 
&   SU(3)&    Octets&
     Meson Cloud\\  \hline\hline
$\frac {\bar{u}}{\bar{d}}$(p),$\frac {\bar{u}}{\bar{s}}(\Sigma^+)$
&$ \frac{1}{2} (\frac{2}{3})$&$\frac {1}{2} (\frac{2}{3})$& 0.29   & $\sim 
\frac {1}{2} $ \\
$\frac{\bar{s}}{\bar{u} + \bar{d}}$$(p)$, $\frac{\bar{d}}{\bar{u} + \bar{s}}$$
(\Sigma^+)$ 
&$\sim \frac{1}{4}$&$\sim \frac{1}{4}$ & 0.42    &$\sim 0.1 $\\
$\frac {\bar{u}}{\bar{d}}(\Sigma^+) $ & ?&$\sim \frac{4}{3}$& 0.54&$ \sim 0.3
$\\
\end{tabular}
\end{table}

    Deviations from SU(3) symmetry can also be expected in the distribution 
function of valence quarks \cite{Alb}. For instance, on the basis of a quark-
diquark model, we predict that $\frac {s}{u} (\Sigma^+)$ is more
than three times as large as the SU(3) value at $x \sim 0.7$.

    Appropriate for this conference in honor of Josef Speth's 60th birthday,
we have used the Sullivan process to compute the valence and sea quark
distribution functions in the $\Sigma^+$. We have 

\begin{equation}
\Sigma^+ = \sqrt{Z}[\Sigma^+_{bare} + \sum\int dy d^2k_\perp \phi_{BM}
(y, k_\perp^2) B (y, \vec{k}_\perp) M(1-y, - \vec{k}_\perp) ] \qquad,
\end{equation}
 with $M = \pi^+, \pi^0, \bar{K}^0 $ and $ B = \Lambda^0, \Sigma^0, 
\Sigma^+, p$.

    We carry out our calculation in the infinite momentum frame with time 
ordered perturbation theory \cite {Sp}
and pseudoscalar coupling. We neglect masses above 
1700 MeV and thus do not consider $\Delta \bar{K}$ states. We have respected
the necessary symmetries. \cite{Sp}For instance, we have
\begin{equation}
q(\Sigma^+, x) = \sqrt{Z} (q_{bare} + \delta q) \qquad ,
\end{equation}
with
\begin{equation}
\delta q (\Sigma^+, x) = \sum \int_x^1 f_{MB}(y) q_M(\frac{x}{y})\frac{dy}{y}
+ \int_x^1 f_{BM}(y) q_B(\frac{x}{y}) \frac{dy}{y} \qquad,
\end{equation}
and require 

\begin{equation}
f_{MB}(y) = f_{BM}(y)
\end{equation}

In order to take finite sizes into account, we introduce Gaussian form factors
with the size set by $\Lambda = 1.08 GeV$ \cite{Sp}. We have studied the 
dependence of our results on $\Lambda$; the changes are quantitative, but not
qualitative. Coupling constants are taken from 
Dumbrajs, Koch, and Pilkhun. \cite {DKP} We assume that 20\% of the mesons' 
momenta are carried by sea quarks. 

   For the $s$ quark distribution in the $\bar {K}^0, 
\Lambda^0$, and $\Sigma $, we
take both SU(3) and a shifted distribution which takes the higher mass of the 
$s$ quark into account. 

    For the proton, we find that our calculation with the omission of 
the $\Delta$ and higher mass (e.g., vector) mesons gives an acceptable fits
for the $ (\bar{d} - \bar{u})$ experimental data.\cite {E866} However,
$ \bar{d} / \bar{u} $ shows no decrease at higher values of $x$,
contrary to experiment; see, however \cite {Mel}

    The results of our calculation are shown in the following figures.
Fig. 1 shows the momentum fraction carried by $u$ and $\bar{u}$ quarks; 
Figs. 2 and 3 are 
the same for $s$ and $d$ quarks; Figs. 4 and 5 show the valence quark momentum
distributions. Figs. 6 and 7 show $\bar{d}/\bar{u}$ and ($ \bar{d} - \bar{u}$)
distributions. Fig. 8 is the momentum fraction $ x( \bar{d} - \bar{u})$. 
Figs. 9 and 10 compare the $\Sigma^+$ and proton $\bar{r}\equiv \bar{d} / 
\bar{u}$ and $(\bar{d} - \bar{u})$. It is readily apparent that
$(\bar{d} - \bar{u})$ is larger in the $\Sigma^+$ than in the proton. The 
ratio $\bar{r} \equiv  \bar{d} /\bar{u}$ in the $\Sigma^+$ vs. $p$ is seen 
to begin at approximately 1 at small $x$ and to climb to 2 at $x \sim 0.35 $.

    In conclusion, the ratio $ \bar{r} \equiv \bar{d} /\bar{u}$ in the
$\Sigma^+$ may be $<$ 1, as in (SU(3) or $>$ 1 as in the meson cloud model. 
We have calculated both $q(x)$ and $ \bar{q}(x)$ in the meson cloud model and 
have confirmed that $\bar {r}(\Sigma^+) > \bar {r}(p)$ in this model.

\listoffigures

Fig. 1. Momentum fraction carried by sea $u$ and $\bar{u}$ quarks.
        The dashed curve here and elsewhere corresponds to the bare sea.

Fig. 2. Momentum fraction carried by sea $s$ and $\bar{s}$ quarks.

Fig. 3. Momentum fraction carried by sea $d$ quarks. The figure for $\bar{d}$
        quarks is similar.

Fig. 4. Momentum fraction carried by valence $u$ quarks.

Fig. 5. Momentum fraction carried by valence $s$ quarks.

Fig. 6. The distribution of the ratio $\bar{d} / \bar{u}$.

Fig. 7. The distribution ($\bar{d} - \bar{u}$) for the $\Sigma^+$.

Fig. 8. The momentum fraction $ x$ ($\bar{d} - \bar{u}$).

Fig. 9. The ratio $\bar{r}(\Sigma^+) / \bar{r}(p)$.

Fig. 10.The ratio $(\bar{d} - \bar{u})(\Sigma^+) / (\bar{d} - \bar{u})(p)$.
\end{document}